\documentclass[twoside]{article}
\usepackage[a4paper]{geometry}

\usepackage[utf8]{inputenc}
\usepackage[OT1]{fontenc}

\usepackage{RR}
\usepackage{hyperref}

\usepackage{fleqn}

\usepackage{amsmath}
\usepackage[psamsfonts]{amssymb}
\usepackage{MnSymbol}

\usepackage{color}
\usepackage{url}

\usepackage{xspace}
\usepackage{makeidx}
\usepackage{titletoc}

\usepackage{graphicx}

\usepackage{pgf}
\usepackage{tikz}
\usetikzlibrary{trees}
\usetikzlibrary{matrix}

\usepackage{comment}

\includecomment{RR}\excludecomment{IEICE}

\newif\ifshowlabel\showlabelfalse

\usepackage{listings} 

\definecolor{mygreen}{rgb}{0,0.4,0}
\definecolor{stringcolor}{rgb}{0,0.4,0}

\definecolor{mygray}{rgb}{0.5,0.5,0.5}
\definecolor{commentcolor}{rgb}{0.5,0.5,0.5}

\definecolor{mymauve}{rgb}{0.58,0,0.82}
\definecolor{lexercolor}{rgb}{0.58,0,0.82}

\definecolor{myblue}{rgb}{0,0,0.6}
\definecolor{actioncolor}{rgb}{0,0,0.6}

\definecolor{myblue2}{rgb}{0,0,0.6}
\definecolor{attributecolor}{rgb}{0,0,0.7}

\definecolor{mydeepblue}{rgb}{0,0,0.4}
\definecolor{bytecode}{rgb}{0.2,0.1,0.61}
\definecolor{symbolcolor}{rgb}{0,0,0.4}
\definecolor{fctcolor}{rgb}{0,0,0.4}

\definecolor{eventcolor}{rgb}{1,0,0}

\lstdefinelanguage{antescofo}{
  columns=fixed, %% flexible, spaceflexible, fullflexible
  basicstyle=\small\tt,
  resetmargins=true,
  xleftmargin=-6mm,
  morecomment=[l]{//},
  morecomment=[l]{;},
  morecomment=[s]{/*}{*/},
  commentstyle=\color{commentcolor}, 
  morestring=[b][\color{stringcolor}]",
  sensitive=false,
  morekeywords=[1]{ 
    @sin, @asin, @sinh, @cos, @acos, @cosh, @tan, @atan, @exp, 
    @log10, @log2, @log, @pow, @sqrt, @rand, @abs, @size, @min, @max,
    @floor, @ceil,
    @date, @rdate,
    @map_history, @map_history_date, @map_history_rdate,
    @make_score_map, @make_duration_map,
    @is_vector, @is_integer_indexed, @is_list, @is_defined, 
    @is_int, @is_float, @is_numeric, @is_bool, @is_string, @is_symbol, 
    @is_map, @is_function, @is_interpolatedmap, @is_undef, @is_fct, 
    @select_map, @shift_map, @gshift_map, @map_val, @merge, @add_pair,
    @min_key, @max_key, @min_val, @max_val,
    @history_map, @history_map_date, @history_map_rdate,
    @listify, @map_reverse, @compose_map,
    @tab_map,@concat,@tab_reverse, @push_back
  },
  keywordstyle=[1]{\color{fctcolor}},
  otherkeywords={
    antescofo::actions, 
    antescofo::analysis, 
    antescofo::antescofo\_tempo, 
    antescofo::before\_nextlabel, 
    antescofo::bpmtolerance,
    antescofo::calibrate, 
    antescofo::clear, 
    antescofo::gamma, 
    antescofo::getcues, 
    antescofo::getlabels, 
    antescofo::gotobeat, 
    antescofo::gotocue,
    antescofo::gotolabel, 
    antescofo::harmlist, 
    antescofo::info, 
    antescofo::jumptocue, 
    antescofo::jumptolabel, 
    antescofo::killall, 
    antescofo::mode,
    antescofo::nextaction, 
    antescofo::nextevent, 
    antescofo::nextfwd, 
    antescofo::nextlabel, 
    antescofo::nofharm, 
    antescofo::normin, 
    antescofo::obsexp,
    antescofo::pedalcoeff, 
    antescofo::pedaltime, 
    antescofo::pedal, 
    antescofo::piano, 
    antescofo::playfrombeat, 
    antescofo::playfrom, 
    antescofo::play,
    antescofo::preload, 
    antescofo::preventzigzag, 
    antescofo::previousevent, 
    antescofo::previouslabel, 
    antescofo::printfwd,
    antescofo::printscore, 
    antescofo::read, 
    antescofo::report, 
    antescofo::score, 
    antescofo::start, 
    antescofo::stop, 
    antescofo::suivi, 
    antescofo::tempo,
    antescofo::tempoinit,
    antescofo::temposmoothness, 
    antescofo::tune, 
    antescofo::variance, 
    antescofo::verbosity, 
    antescofo::verify, 
    antescofo::version
  },
  keywordstyle={\color{fctcolor}},
  morekeywords=[2]{  
    chord,
    event,
    @hook,
    jump,
    @jump,
    @modulate,
    multi,
    note,
    trill
  },
  keywordstyle=[2]{\color{eventcolor}},
  morekeywords=[3]{
    napro_trace,
    bind,
    bpm,
    inlet,
    off,
    on,
    tempo,
    @tempo,
    transpose,
    @transpose,
    variance
  },
  keywordstyle=[3]{\color{eventcolor}},
  morekeywords=[4]{ 
    @LID, @lid,
    @UID, @uid,
    @fun_def,
    @insert, 
    @macro_def,
    @proc_def
  },
  keywordstyle=[4]{\color{lexercolor}},
  morekeywords=[5]{
    abort,
    action,
    assert,
    cfwd,
    closefile,
    curve,
    do,
    else,
    expr,
    false,
    gfwd,
    glob,
    global,
    group,
    if,
    in,
    imap,
    kill,
    let,
    lfwd,
    loc,
    local,
    loop,
    map,
    ms,
    of,
    openoutfile,
    oscoff,
    oscon,
    oscrecv,
    oscsend,
    parfor,
    port,
    s,
    symb,
    tab,
    true,
    type,
    until,
    when,
    whenever,
    while
  },
  keywordstyle=[5]{\color{actioncolor}},
  morekeywords=[6]{ 
    @action,
    @coef,
    @global, 
    @grain,
    @label,
    @local,
    @name,
    @norec,
    @tempo,
    @tight,
    @type
  },
  keywordstyle=[6]{\color{attributecolor}}
}

\newcommand{\code}[1]{\lstinline;#1;}

\def\<#1>{\langle #1 \rangle}
\newcommand{\B}{\mathcal{B}\xspace}

\newcommand{\In}{\mathcal{I}\xspace}
\newcommand{\Out}{\mathcal{O}\xspace}
\newcommand{\M}{\mathcal{M}\xspace}
\newcommand{\Q}{\mathcal{Q}\xspace}

\newcommand{\X}{\mathcal{X}\xspace}
\newcommand{\XG}{{\X_\mathit{g}}\xspace}
\newcommand{\XL}{{\X_\mathit{l}}\xspace}

\newcommand{\nexte}{\mathit{next}\xspace}

\newcommand{\dom}{\mathit{dom}}
\newcommand{\true}{\mathit{true}}
\newcommand{\false}{\mathit{false}}
\newcommand{\TRUE}{\textsc{true}}
\newcommand{\FALSE}{\textsc{false}}
\newcommand{\XOR}{\textsc{xor}}
\newcommand{\SOR}{\textsc{sor}}
\newcommand{\AND}{\textsc{and}}
\newcommand{\ERROR}{\textsc{error}}

\lstdefinelanguage{ascoc}{
  columns=fixed, %% flexible, spaceflexible, fullflexible
  basicstyle=\small\tt,
  resetmargins=true,
  xleftmargin=-6mm,  
  morecomment=[l]{//},
  morecomment=[s]{/*}{*/},
  keywordstyle=\color{fctcolor}\bfseries
  commentstyle=\color{commentcolor}, 
  morestring=[b][\color{stringcolor}]",
  sensitive=false,
  morekeywords=[1]{ 
    emit, send, :=, jump, if, then, else, stop, spawn, read,
    receive, jump, 
    present, await, suspend, 
    repeat, for,
    switch, asap, sustain},
  keywordstyle=[1]{\color{bytecode}}
}  

\newcommand{\ascode}[1]{\lstinline;#1;}

\newcommand{\emit}{\lstinline[language=ascoc]!emit!\xspace}
\newcommand{\send}{\lstinline[language=ascoc]!send!\xspace}
\newcommand{\cond}{\lstinline[language=ascoc]!if!\xspace}

\newcommand{\jump}{\mathrel{\lstinline[language=ascoc]!jump!}}
\newcommand{\assign}{\mathrel{\lstinline[language=ascoc]!:=!}}
\newcommand{\halt}{\lstinline[language=ascoc]!stop!\xspace}
\newcommand{\switch}{\lstinline[language=ascoc]!asap!\xspace}
\newcommand{\sustain}{\lstinline[language=ascoc]!sustain!\xspace}

\newcommand{\await}{\lstinline[language=ascoc]!await!\xspace}
\newcommand{\wait}{\lstinline[language=ascoc]!wait!\xspace}
\newcommand{\receive}{\lstinline[language=ascoc]!receive!\xspace}
\newcommand{\present}{\lstinline[language=ascoc]!present!\xspace}

\newcommand{\suspend}{\lstinline[language=ascoc]!suspend!\xspace}

\newcommand{\spawn}{\lstinline[language=ascoc]!spawn!\xspace}

\newcommand{\iteration}{\lstinline[language=ascoc]!repeat!\xspace}
\newcommand{\fordur}{\lstinline[language=ascoc]!for!\xspace}

\RRNo{8520}
\RRdate{April 2014}
\RRauthor{% les auteurs
Florent Jacquemard\thanks{INRIA}%
  \thanks[sfn]{Ircam, UMR STMS 9912 CNRS/UPMC}%
\and
Clément Poncelet\thanks{DGA \& INRIA}%
\thanksref{sfn}}
\authorhead{F.~Jacquemard \& C.~Poncelet}
\RRtitle{Représentation Intermédiaire pour le langage d'Antescofo}
\RRetitle{Antescofo\\ Intermediate Representation}
\titlehead{Antescofo Intermediate Representation}
\RRnote{This work has been supported by the ANR project Inedit (ANR-12-CORD-009) \url{http:// inedit.ircam.fr}.}
\RRresume{Ce rapport décrit un langage intermédiaire conçu pour la représentation interne 
de programmes du système musical interactif Antescofo.
Il est actuellement utilisé dans des tâches de vérification portant en particulier sur les durées, 
de test de conformité fondé sur modèles, d'analyse statique et de simulation.}

\RRabstract{
We describe an intermediate language 
designed as a medium-level internal representation of programs of the interactive music system Antescofo.
This representation is independent both of the Antescofo source language and of the architecture of the execution platform.
It is used in tasks such as verification of timings, model-based conformance testing, 
static %data-flow 
control-flow analysis or simulation.

This language is essentially a flat representation of Antescofo's code, 
as a finite state machine extended with local and global variables, 
with delays and with concurrent threads creation.
It features a small number of simple instructions
which are either blocking 
(wait for external event, signal or duration) or not 
(variable assignment, message emission and control).
}

\RRmotcle{Systèmes musicaux interactifs, compilation, modèles formels}
\RRkeyword{Interactive Music Systems, Compilation, Formal Models}

\RRprojets{MuTant}

\RCParis
\begin{document}
%%\title{Antescofo Intermediate Code}
\makeRR

%\author{Florent Jacquemard, Clément Poncelet\\
%\small INRIA \& IRCAM UMR STMS 9912 CNRS/UPMC}

%\tableofcontents
%\pagebreak

\section{Intermediate Code: Syntax}
We describe in this section an abstract syntax
for the intermediate code which will be the result of a front-end compilation of Antescofo's programs.
It is defined independently of Antescofo's source language 
and of the architecture of the execution platform.
We give in the description some examples corresponding to the compilation of programs in Antescofo language.

\subsection{Values}

\subsubsection{Atomic Values} \label{sec:scalar}
We assume the same scalar values as in Antescofo, see~\cite{Antescofo13,INRIARR-8422}:
Booleans values $\true$ and $\false$,
the integers, 
the floats (double),
the strings and one 
%keywords (reserved constants) : ...
undefined value (which is not used in this document).
We also assume compounds values for vectors and maps.

Durations are a specific type of value.
They can be expressed with different time units, corresponding to different clocks.
For instance, the seconds is the time unit of the wall clock (physical time).
In Antescofo, the most important time unit is beats, which refers to an inferred tempo.

As explained in Section~\ref{sec:multiclock}, we assume that 
time units are inter-convertible and hence we shall sometimes drop them 
in the expression of delays in the following.

\subsubsection{Variables}
Let $\XG$ and $\XL$ be two disjoint infinite sets
of respectively \emph{global variables} and \emph{local variables}.

\subsubsection{Expressions}
The expressions are the same as in the Antescofo language~\cite{Antescofo13}.
Note that predicates operating on duration values 
will rely on multi clock services described in Section~\ref{sec:multiclock}.
An expression is called \emph{ground}
when it does not contain variables.

\subsection{Symbols}

\subsubsection{Input Symbols}
We assume a given set of input symbols $\In = \{ i_0,\ldots \}$,
called \emph{input events}, 
representing some information %in a finite domain 
expected from the external environment.
%\noindent
The set $\In$ is assumed totally ordered by a function called $\nexte$.
%Moreover, some elements of $\In$ can carry information in infinite domains, 
%and we shall use patterns of the form $i(x)$ where $x \in \XG$ 
%for the assignment of global variables during reception.

We take for instance a set
of Antescofo's \emph{events} (notes etc), 
as defined in Section~2 of~\cite{Antescofo13}, 
together with their positions in the score.
For such an input symbol $i$ at position $n$, 
$\nexte(i)$ is then defined as the event at position $n+1$.

A generalization of the total ordering on input events into 
a DFA with state set $\Q$ and input alphabet $\In$
will be the subject of further work.

\subsubsection{Output Symbols}
We assume a given set of symbols $\Out = \{ a_0, a_1\ldots \}$,
representing action emitted or messages sent to the external environment.

For Antescofo, the elements of $\Out$ are called 
\emph{internal} (atomic) \emph{actions} and can be messages to MAX/MSP, 
OSC messages...

\subsubsection{Signals}
We consider internal \emph{signals} represented by natural numbers, and denoted $s$...

%\emph{Signals} are internal symbols and can have one of the types:
%\begin{itemize}
%\item \emph{basic signals}, represented by natural numbers, denoted $s$,
%\item \emph{valued signals}, denoted $s(e)$, given by a signal number $s$ and an ground expression $e$,
%\item \emph{signal patterns}, denoted $s(x)$ where $s$ is signal number $s$ and $x$ is a variable  in $\XG \uplus \XL$.
%\end{itemize}
%We assume that the set of numbers used for basic signals is disjoint from the set of 
%numbers used for valued signals or signal patterns.  
%
%The valued signals will be used for the emission of signals in the instructions below, 
%while signal patterns which will be used for the reception of signals.

In the case of Antescofo, typical signals include 
the name of groups, kill signals and signals associated to missed events
(similar to exceptions).

%We also consider \emph{signals expressions} which are Boolean combinations of signals.
%Given a set $\alpha$ of signals and a signal expression $b$, 
%we write $\alpha \models b$ to express that $b$ evaluates to true 
%when every signal of $\alpha$ is replaced by true in $b$
%and every other signal is replaced by false in $b$.

\subsection{Machines} \label{sec:machine}
A \emph{machine} $\M$ is an table of fixed size
containing instructions in the set presented below.
A \emph{location} $\ell$ is an index in the table (natural number).
We assume a fixed total ordering $\ll$ on locations of $\M$. 
It will be used to reflect the order of instructions in the source Antescofo program.
Therefore, the ordering $\ll$ may differ from the ordering on natural numbers.
However, for the sake of readability, we write $\ell+1$ for the 
the successor of $\ell$ wrt $\ll$.

\subsection{Instructions}
We now enumerate the instructions of the intermediate code, 
with informal descriptions 
(Section~\ref{sec:sem} provides a detailed definition of semantics).

Every instruction has an implicit \emph{source location} $\ell$
which is its index in the table $\M$.
It can have zero, one or several \emph{target location} denoted $\ell'$...

We consider two categories of instructions. 
The \emph{synchronous} instructions are instantaneous: 
they are executed simultaneously, in a single logical instant.
The \emph{asynchronous} instructions are blocking: 
they stop the computation, waiting for an event to happen.
Time is flowing while waiting during the execution of an asynchronous instruction $\kappa$, 
and the date of the event unlocking $\kappa$ defines a new logical instant,
as explained in Section~\ref{sec:sem-time}.

\subsubsection{Atomic Synchronous Instructions} 
All these instructions are executed within the same logical instant.
%Each of them has a unique target location $q' = q+1$
%(the next location in the table), which is written explicitly for technical convenience, 
%except $\halt$ which has no target location.

\begin{description}
\item $\emit\, s$, where $s$ is an internal signal.
Signal emission: broadcast the signal $s$,
and continue at $\ell + 1$ with the next instruction.

%\item $\emit\, s(e) \to q'$, where $s$ is a valued signal. 
%Valued signal emission:
%broadcast the valued signal $s$, where $e$ is an expression, and continue in $q'$.

\item $\send\, a$, where $a \in \Out$ is an output symbol.
Message sending: send $a$ to the external environment (e.g. OSC or MAX message), 
and continue at $\ell + 1$ with the next instruction.

\item $x\assign e$, where $x$ is a local or global variable.
Variable assignment.
%The instruction ...

\item $\halt$.
Terminates the execution.
\end{description}

\subsubsection{Branching Synchronous Instruction} 
%This instruction has one target locations, $q' = q+1$ and $q''$.

\begin{description}
\item $\cond e \jump \ell'$.
Conditional:
if the Boolean expression $e$ evaluates to true, then jump to the location $\ell'$, 
otherwise continue at $\ell+1$ with the next instruction.
\end{description}

\subsubsection{Concurrent Synchronous Instructions} 
The two following instructions start a concurrent execution, 
with passing or not of the local environment.

\begin{description}
\item $\spawn\, \ell'$.
Continue the current thread with the next instruction at $\ell+1$,
and start concurrently a new thread at location $\ell'$ with a copy 
of the local environment.

\item $\spawn_0\, \ell'$.
Continue the current thread with the next instruction at $\ell+1$,
and start concurrently a new thread at location $\ell'$ 
with an new empty local environment.
\end{description}

\subsubsection{Atomic Asynchronous Instructions} 
The following instructions let the time flow.
Each of them has an explicit target location $\ell'$.

\begin{description}
\item $\await\, e \jump \ell'$,
where $e$ is an expression that must be evaluable in a duration value $d$ 
in a time unit $\mathit{tu}$.
Wait for $d$ units of the time units $\mathit{tu}$,
and jump to location $\ell'$.

\item[(opt)] $\iteration\, e \jump \ell' \fordur\, e'$,
where $e$ and $e'$ are expressions that must be evaluable 
in duration values $d$ and $d'$
in respective time units $\mathit{tu}$ and $\mathit{tu}'$.
Periodically wait for $d$ units of the time unit $\mathit{tu}$,
and at each iteration, 
create a new thread at location $\ell'$.
Stop iterating after $d'$ units of the time unit $\mathit{tu}'$.

\item $\receive\, i \jump \ell'$, where $i \in \In$ is an input event.
Wait for the reception of the input event $i$ and jump to location~$\ell'$.

\item $\present\, s \jump \ell'$, where $s$ is a signal.
Wait for $s$ and jump to location~$\ell'$.

%\item $\present\, s(x) \to q'$,
%where $s$ is a signal number and $x$ is a local variable in $\XL$.
%Wait for the signal $s$, 
%assign the value received along with the signal $s$ to $x$
%and jump to location $q'$.

\item $\suspend\, e \jump \ell'$, where $e$ is a boolean expression.
Wait for $e$ to become evaluable to true and then jump to location~$\ell'$.
\end{description}

%\paragraph{Compound instructions.}
%The instruction $\waitemit\; d\; i$,
%and the instruction $\waitsend\; d\; m$...

Note the difference between the synchronous $\cond$
and the asynchronous $\suspend$: 
The former evaluates immediately the associated expression
(with failure when it is not evaluable)
whereas the latter blocking instruction waits until the expression is evaluable to true.

The instruction $\iteration$ can be encoded using 
a combination of $\sustain$, $\await$ and $\spawn_0$, 
see Figure~\ref{fig:iteration}, but it is 
more efficient to use this instruction
which rely on a special clock service described in 
Section~\ref{sec:multiclock}, 
and avoids to start a timer at each iteration.

\begin{figure}
\begin{center}
\begin{tabular}{ccc}
\begin{tabular}[t]{ll}
$\ell$   & $\sustain\; (\ell'-3)\, (\ell'-1)$\\
$\ell+1$ & next instruction\\
\vdots
\end{tabular}
&
\begin{tabular}[t]{ll}
$\ell'-3$   & $\spawn_0\; \ell'$\\
$\ell'-2$ & $\await\, e \jump \ell'-3$\\
$\ell'-1$ & $\await\, e' \jump \ell'+k$\\
\end{tabular}
&
\begin{tabular}[t]{ll}
$\ell'$ & code of the loop\\
\vdots\\
$\ell'+k$ & $\halt$
\end{tabular}
\end{tabular}
\end{center}
%\begin{lstlisting}
%\end{lstlisting}
\label{fig:iteration}
\caption{Encoding of $\iteration\, e \jump \ell' \fordur\, e'$}
\end{figure}

\subsubsection{Branching Asynchronous Instruction} 
\begin{description}
\item $\switch\; L$, where $L$ is a non-empty list $\ell_1 \ldots \ell_n$ 
of locations of asynchronous transitions in $\M$.
Wait concurrently (competitively) for the atomic asynchronous instructions
$\M(\ell_1)$, \ldots $\M(\ell_n)$. 
Once one instruction $\M(\ell_i)$ in unlocked, jump to its target. 
The other instructions are discarded.
%The target locations are the target locations of the individual instructions in $K$.
%
%The set $K$ must contain at most one $\await$ instruction, 
%and either one of the following:
%\begin{itemize}
%\item at most one $\iread$ instruction, 
%\item at most one $\suspend$ instruction, 
%\item at most one $\present\, s(x)$ instruction, 
%\item possibly several instructions of the form $\present\, b_1$, \ldots, $\present\, b_n$
%      such that the Boolean combination of signals $b_1$,\ldots, $b_n$ are mutually exclusive. 
%\end{itemize}
\item $\sustain\; \ell_1\, \ell_2$, 
where $\ell_2$ is the location of an asynchronous instruction in $\M$.
Every asynchronous instruction $\kappa$ following $\ell_1$ 
will be controlled by $\M(\ell_2)$.
If $\kappa$ is unlocked before $\M(\ell_2)$, 
then the execution continues at the target of $\kappa$,
but $\M(\ell_2)$ is not discarded (unlike with $\switch$).
If $\M(\ell_2)$ is unlocked before $\kappa$, jump to its target
and $\kappa$ is discarded.
\end{description}

The instruction $\sustain$ could be encoded by
adding $\ell_2$ (in $\switch$ instructions)
to every asynchronous instruction following $\ell_1$.
It has been added to lighten notations, 
and is similar to the construction of hierarchical states
(see e.g. \cite{Harel87statecharts}, \cite{LeeEtAl14Ptolemy}, \cite{Ghosal06HTL}).

\section{Intermediate Code: Semantics} \label{sec:sem}
We present in this section the execution of a machine $\M$.
It follows reactive synchronous semantics, with concurrent thread creation and 
cooperative multitasking.
It extends previous works on the timed-automata based
definition of an operational semantics of the static kernel of Antescofo~\cite{Echeveste13jdeds}.

Intuitively, the machine $\M$ is ran by several concurrent "threads", 
organized in a tree structure (called \emph{global tree}).
Each thread (called \emph{local state} in Section~\ref{sec:state}) 
points to a line $\ell$ in $\M$. % and has a \emph{store} $\sigma$.
There is also a global store $\gamma$, for assignment of global (shared) variables, not attached to a particular thread.
One step of execution of $M$, at instant $t_k$ consists in the following successive steps.
\begin{enumerate}
\item For every thread, iteratively execute the pointed instruction
      as long as it is \emph{synchronous}.
The order of execution is defined after $\ll$ (see Section~\ref{sec:machine}).
The executions are assumed instantaneous  
(hypothesis of \emph{synchronicity}): the date is still $t_k$ during the 
execution of all successive synchronous instructions.
When done (\textit{i.e.} after step 1 and before step 2)
every thread points to an asynchronous instruction.
\item Wait, during a delay $d$, for a \emph{logical event}, which can be
%that will unlock at least one of the pointed asynchronous instructions:
\begin{itemize}
\item a signal sent or a global variable modified during step 1 (in this case $d = 0$)
\item an external input event
\item an external modification of a global variable % (unlocking a $\suspend$)
\item the expiration of a delay (following an instruction $\await$ or $\iteration$).
\end{itemize}
Then execute the unlocked (asynchronous) instructions (do $\jump$'s)
and reorganize the global thread tree.
\item This defines a new logical instant $t_{k+1} = t_k + d$.
Restart 1.
\end{enumerate}

\subsection{Multiclock Services} \label{sec:multiclock}
Several instructions explicitly refer to duration values.
As explained in Section~\ref{sec:scalar}, 
durations values can be expressed with different time units, 
corresponding to different clocks.
We assume that these clocks are managed in an external module (called \emph{clocks module})
accessible through services described as follows.

\begin{itemize}
\item it is possible to be notified at any time of the current date in any time unit
      (it is needed for dealing with some reserved variables in expressions).
\item any two delays in same or different units are comparable
      (it is needed for evaluating some Boolean predicates on durations in the expressions).
\item \label{clock-service:timer}
it is possible to start a \emph{timer} attached to a node $p$ in the global thread tree,
given a delay $d$ in a time unit $\mathit{tu}$.

The node $p$ will be notified of the expiration of the delay 
after $d$ units of~$\mathit{tu}$.
\item[(opt)] \label{clock-service:timetrigger}
it is possible to start a \emph{recursive timer}
given a period value $d$ in a time unit $\mathit{tu}$, 
a delay $d'$ in a time unit $\mathit{tu}'$, 
and a node $p$ in the global thread tree.

The node $p$ will be notified every $d$ units of $\mathit{tu}$
until the expiration of the delay $d'$.
\end{itemize}
The notifications of expiration are considered in the same way as external events in Section~\ref{sec:sem-asynchronous}.

\noindent
The recursive timers are used to represent directly Antescofo's periodic loops 
with an expiration date.

%inter-convertibility.
%These properties are ensured in particular when 
%every time unit $\mathit{tu}$ corresponds to a function 
%%$f_\mathit{tu}: 
%from $\mathbb{R}_+$ to $\mathbb{R}_+$ which converts a 
%date expressed in $\mathit{tu}$ into a date expressed in seconds.
%These functions are not specified, they are only assumed to 
%be continuous, strictly increasing and divergent (to avoid Zeno behavior).

%We might compare any two durations values expressed in different time units 
%through their respective conversion function
%(comparing the corresponding values in seconds).

\subsection{States} \label{sec:state}
A \emph{local store} 
is a mapping from a finite subset of $\XL$ into values.
Given a local store $\sigma$, $x\in \XL$ and a value $v$, 
we write $\sigma[x \mapsto v]$ the store $\sigma'$ defined 
by 
$\dom(\sigma') = \dom(\sigma) \cup \{ x\}$
and $\sigma'(x) = v$ 
and $\sigma'(y) = \sigma(y)$ for all $y \in \dom(\sigma) \setminus \{ x \}$.

\noindent
A \emph{global store} 
is a mapping from a finite subset of $\XG$ into values
and from the finite set of signals occurring in $\M$ into Boolean values.
The latter part is used to accumulate signals sent during the execution of synchronous instructions.

We shall use a similar notation for global stores and local stores.
By abuse of notation, we make no distinction between a store and
his homomorphic extension to expressions.
%associates to every variable see~\cite{Antescofo13}

\noindent
A \emph{local state} is a pair denoted $\< \ell, \sigma>$
where $\ell$ is a location instruction and $\sigma$ is a local store.
It is called synchronous when $\M(\ell)$ is a synchronous instruction, 
and asynchronous when $\M(\ell)$ is an asynchronous instruction.

\noindent
A concurrent state expression $T$, or \emph{tree} for short, 
is either %the smallest set containing
a local state
or one of 
$\TRUE$, $\FALSE$, $\ERROR$,
$\AND(T_1, T_2)$, 
$\XOR(T_1, T_2)$, 
$\SOR(T_1, T_2)$, 
where $T_1$ and $T_2$ are trees.
The operators $\AND$ and $\XOR$ are associative and commutative
(not $\SOR$).
We use the notation $C[T_1]$ to denote a tree made 
of a context $C$ and a subtree $T_1$.
The evaluation of the trees is defined in Section~\ref{sec:sem-asynchronous}.

\noindent
The \emph{global state} is a pair $\< \gamma, T>$
where $\gamma$ is a global store
and $T$ is a tree called \emph{global tree}.
%It is called asynchronous when $L$ contains only asynchronous local states and
%synchronous otherwise.
%We use the operator $\cat$ for the concatenation of sequences.

\subsection{Synchronous Transitions} \label{sec:sem-synchronous}
A synchronous transition between global states represent a maximal execution 
of successive synchronous instructions, 
until the global state contains only asynchronous instructions.
The synchronous instructions are executed sequentially, 
following the ordering $\ll$.
%Intuitively, the execution starts with the first synchronous local state in $g$.
%This local state in $g$ represents a thread executing all synchronous instructions until an asynchronous instruction is met.
%Then, the control is passed to the next synchronous local state in $g$ and so on, 
%until all local states in $g$ are asynchronous (then an asynchronous transition is performed on the global state obtained).
The signals sent during the execution of synchronous instructions 
are accumulated in the global store. % during the synchronous transitions.

We define synchronous transitions with a small step semantics, 
based on a binary relation, denoted $\to$, on global states, 
representing the execution of one synchronous instruction.
Let $g = \< \gamma, C[ \<\ell, \sigma>] >$ be a global state.
We define the relation $\to$ according to the case of $\M(\ell)$.

\begin{description}
\item 
if $\M(\ell) = \emit\, s$ then
\( g \to \bigl\langle  \gamma[s \mapsto \true], C[\<\ell+1, \sigma>]\bigr\rangle  \)
\item  
if $\M(\ell) = \send\, a$, then
\( g \to \bigl\langle \gamma, C[\<\ell+1, \sigma>]\bigr\rangle  \)
\item 
if $\M(\ell) = x\assign e$, 
$x$ is local 
and $\gamma(\sigma(e))$ evaluates to $v$, then
\( g \to 
  \bigl\langle \gamma, C[\<\ell+1, \sigma[x \mapsto v>]\bigr\rangle  \)
\item 
if $\M(\ell) = x\assign e$, 
$x$ is global 
and $\gamma(\sigma(e))$ evaluates to $v$, then
\( g \to 
  \bigl\langle \gamma[x \mapsto v], C[\<\ell+1, \sigma>]\bigr\rangle  \)
\item  
if $\M(\ell) = \halt$, then
\( g \to \bigl\langle  \gamma, C[\TRUE]\bigr\rangle  \)
\item 
if $\M(\ell) = \cond e \jump \ell'$,
and $\gamma(\sigma(e))$ evaluates to $\true$, then
\( g \to \bigl\langle \gamma, C[\<\ell', \sigma>]\bigr\rangle  \)
\item 
if $\M(\ell) = \cond e \jump \ell'$,
and $\gamma(\sigma(e))$ evaluates to $\false$, then
\( g \to \bigl\langle \gamma, C[\<\ell+1, \sigma>]\bigr\rangle  \)
\item 
if $\M(\ell) = \spawn\,\ell'$, then
\( g \to \bigl\langle  \gamma, C[\AND(\<\ell+1, \sigma>, \<\ell', \sigma>)]\bigr\rangle \)
\item 
if $\M(\ell) = \spawn_0\,\ell'$, then
\( g \to 
  \bigl\langle  \gamma, C[\AND(\<\ell+1, \sigma>, \<\ell', \emptyset>)]\bigr\rangle \)
%
%\item 
%if $\M(\ell)$ is an atomic asynchronous instruction, then 
%\( g \to 
%  \bigl\langle  \gamma, C[\llangle \M(\ell), \sigma\rrangle_0]\bigr\rangle 
%\)
%
\item 
if $\M(\ell) = \switch\; \ell_1 \ldots \ell_n$, then
\( g \to 
  \bigl\langle  \gamma, C[\XOR(\< \ell_1, \sigma>,\ldots, \< \ell_n, \sigma>)]\bigr\rangle 
\)
\item 
if $\M(\ell) = \sustain\, \ell_1 \jump \ell_2$, then
\( g \to 
  \bigl\langle  \gamma, C[\SOR(\<\ell_1, \sigma>, \<\ell_2, \sigma>)]\bigr\rangle 
\)
\end{description}
Some cases that require $\gamma(\sigma(e))$ to be evaluable.
If this condition is not met, then %the whole global tree 
the node $\< \ell, \sigma>$ is reduced to $\ERROR$.

\noindent
Moreover, we assume that the clock module is called when entering, from $g$, a local state $\<\ell, \sigma>$
at node $p$ of the global tree, in the following cases:
\begin{description}
\item when $\M(\ell) = \await\, e \jump \ell'$, 
if $\gamma(\sigma(e))$ evaluates to a delay value $d$,
then, if $d >0$, start a timer with $d$ and $p$.
Otherwise, the whole global tree reduces to $\ERROR$. 
\item when $\M(\ell) = \iteration\, e \jump \ell' \fordur\,e'$, 
if $\gamma(\sigma(e))$ and $\gamma(\sigma(e'))$
evaluate respectively to delay values $d >0$ and $d'$,
then, if $d' > 0$, start a recursive timer with $d$, $d'$ and $p$.
Otherwise, the whole global tree reduces to $\ERROR$. 
\end{description}

\medskip
The reflexive-transitive closure of $\to$ is denoted $\xrightarrow{*}$, 
and the operator $\downarrow_*$
of normalization by $\to$ is defined by (using postfix notation): 
%$\downarrow^*$ is the subrelation of $\xrightarrow{*}$
$g' = g\downarrow_*$ iff $g \xrightarrow{*} g'$ 
and for all $g''$ such that $g' \xrightarrow{*} g''$ then $g'' = g'$. 
Note that if $g' = g\downarrow_*$ then  
all the local states occurring in $g'$ are asynchronous. %, solid and in mode 0.

\subsection{Asynchronous Transitions} \label{sec:sem-asynchronous}
We define now asynchronous transitions between global states.
For this purpose we use the notion of \emph{logical event}, denoted $\tau$..., 
which is one of
\begin{itemize}
\item $\varepsilon$, representing an internal event,
\item a symbol g\ representing a notification of the
      expiration of a delay to a node $p$ in the global tree,
\item[(opt)] the symbol $\mathsf{step}\,p$ representing a notification of the
      expiration of a recursive delay to a node $p$ in the global tree,
\item an input symbol $i \in \In$, representing the recognition of $i$,
\item a global store $\alpha$ of the form $\{ x \mapsto v \}$,
representing the assignment of the global variable $x$ by the external environment.
\end{itemize}
Each of them represent an event which can unlock asynchronous instructions,
and will be used to define our time model in Section~\ref{sec:sem-time}.

\noindent
We first define relations
$\xrightarrow{\tau, \gamma}$
between local states indexed by 
a logical event~$\tau$ and a global store~$\gamma$.
In some cases, the top symbol in the right-hand-side is marked
(underlined) to indicate that it has been evaluated.
This marking will be used below for 
the definition of further transformations for $\XOR$, $\SOR$, $\AND$. 
%In each of the following cases, we consider
%an initial local state $\<\ell, \sigma>$ at note $p$ in a global tree.
\begin{description}
\item if $\M(\ell) = \await\, e \jump \ell'$, then 
\begin{description}
\item
\( 
\<\ell, \sigma>
\xrightarrow{\varepsilon, \gamma} 
\ERROR 
\)
if $\gamma(\sigma(e))$ does not evaluate to a delay value
\item 
\( 
\<\ell, \sigma>
\xrightarrow{\varepsilon, \gamma} 
\underline{\<\ell', \sigma>}
\)
if $\gamma(\sigma(e))$ evaluates to a delay 0
\item
\( 
\<\ell, \sigma>
\xrightarrow{\mathsf{done}\,p, \gamma}    
\underline{\<\ell', \sigma>}
\)
if the local state $\<\ell, \sigma>$ occurs at node $p$ in the global tree
\end{description}
\item if $\M(\ell) = \iteration\, e \jump \ell' \fordur\, e'$, then
\begin{description}
\item
\( 
\<\ell, \sigma>
\xrightarrow{\varepsilon, \gamma} 
\ERROR 
\)
if $\gamma(\sigma(e))$ or $\gamma(\sigma(e'))$ does not evaluate to a delay value,
or if $\gamma(\sigma(e))$ evaluates to a delay 0
\item 
\( 
\<\ell, \sigma>
\xrightarrow{\varepsilon, \gamma} 
\underline{\<\ell', \sigma>}
\)
if $\gamma(\sigma(e'))$ evaluates to a delay 0
\item
\( 
\<\ell, \sigma>
\xrightarrow{\mathsf{step}\,p, \gamma}    
\underline{\AND}(\< \ell, \sigma>, \< \ell', \emptyset>)
\)
if the local state $\<\ell, \sigma>$ occurs at node $p$ in the global tree
\end{description}
\item if $\M(\ell) = \receive\, i \jump \ell'$, then
\begin{description}
\item
\(
\<\ell, \sigma>
\xrightarrow{i, \gamma} 
\underline{\<\ell', \sigma>}
\)
\end{description}
\item if $\M(\ell) = \present\, s \jump \ell'$, then
\begin{description}
\item 
\( 
\<\ell, \sigma>
\xrightarrow{\varepsilon, \gamma}
\underline{\<\ell', \sigma>}
\)
if $\gamma(s) = \true$
\end{description}
\item if $\M(\ell) = \suspend\, e \jump \ell'$, then
\begin{description}
\item 
\( 
\<\ell, \sigma>
\xrightarrow{\varepsilon, \gamma} 
\underline{\<\ell', \sigma>} 
\)
if $\gamma(\sigma(e))$ evaluates to $\true$
\item
\( 
\<\ell, \sigma>
\xrightarrow{\alpha, \gamma} 
\underline{\<\ell', \sigma>}
\)
if $\alpha$ is a global store
and $\alpha(\gamma(\sigma(e)))$ evaluates to $\true$
\end{description}
\end{description}
% explicit.
%The cases where an expression $e$ (for a delay or a condition)
%cannot be evaluated because it contains a variable
%not in the domain of $\sigma$ and $\gamma$ provokes the termination of the execution with an error.

We define below another set of transformation rules for trees called
normalization rules.
In these rules, $T$, $\underline{T}$, $\underline{T'}$, $X$
represent trees that cannot be transformed anymore (normal forms).
Moreover, the top symbol of $\underline{T}$, $\underline{T'}$ is marked, 
the top symbol of $T$ is unmarked, 
and the top symbol of $X$ is either marked or unmarked.
%(respectively unmarked).
Note in particular that $\underline{T}$ cannot be $\ERROR$.
Remember that $\XOR$ and $\AND$ are associative and commutative.
\begin{description}
\item
\(
\XOR(\underline{T}, T)
\to
\underline{T}
\)
%if $T \neq \ERROR$
%and $T$ does not have the form $\<\ell', \sigma'>$
%
\item 
\(
\XOR(\underline{T}, \underline{T'})
\to 
\ERROR
\)
\item 
\(
\XOR(X, \ERROR)
\to
\ERROR
\)
\item 
\(
\SOR(T, \underline{T})
\to
\underline{T}
\)
\item 
\(
\SOR(\underline{T}, \underline{T'})
\to 
\ERROR
\)
\item 
\( \SOR(X, \ERROR) \to \ERROR \),
\( \SOR(\ERROR, X) \to \ERROR \)
\item 
\(
\AND(X, \TRUE)
\to
X
\)
\item 
\(
\AND(X, \ERROR)
\to
\ERROR
\)
\end{description}

\medskip\noindent
We denote $T\downarrow_{\tau, \gamma}$ 
the tree $T'$ obtained from $T$ in three steps: 
\begin{enumerate}
\item application of 
      the rules $\xrightarrow{\tau, \gamma}$      
      at most once to each leaf of $T$.
      When $\xrightarrow{\tau, \gamma}$ is appliable
      to one leaf at least, then we say that 
      the logical event $\tau$ \emph{unlocks} $T$ 
      \emph{wrt} the global store $\gamma$.
\item iterated application of the normalization rules
      to internal nodes, as long as possible
\item finally, removing of the marks
(i.e. $\underline{\<\ell', \sigma>}$ is renamed into $\<\ell', \sigma>$,
 $\underline{\AND}$ is renamed into $\AND$ etc).
\end{enumerate}

%Let $T$ be a global tree containing only solid local states in mode $0$.
%We say that a logical event $\tau$ \emph{unlocks} $T$ wrt a global store $\gamma$
%if $T' = T\downarrow_{\tau, \gamma}$ is $\TRUE$ or it 
%contains at least one non-solid local state.

%\subsection{Model of Time} 
\subsection{Execution}  \label{sec:sem-time}
The execution of a machine $\M$ is a sequence of global states, 
each of them being obtained from the previous one in two steps: 
one synchronous transition 
(defined in Section~\ref{sec:sem-synchronous} as a maximal sequence 
 of execution of synchronous instructions),
followed by one asynchronous transition
(defined in Section~\ref{sec:sem-asynchronous}
 as the parallel and simultaneous execution of asynchronous transitions).
The dates of appearance of each global state 
(the beginning of execution of synchronous transitions) 
will be called \emph{logical instants}.
They correspond to the dates of logical events
described at the beginning of Section~\ref{sec:sem-asynchronous}. 
Following the time model of Antescofo (see~\S~3 of \cite{Antescofo13}).
Hence every new logical instant correspond to one of:
the expiration of a delay, 
the recognition of an input event or internal signal,
the assignment of a global variable by the external environment.
% (e.g. through an OSC message or a MAX/PD binding)

%The last kind of instant can cause the modification of the global store
%(the global store can also be modified by assignment instructions as seen in Section~\ref{sec:sem-synchronous}).

%every signal sent is valid only during the next step

Formally, let us define the first logical instant as $t_0 = 0$
and assume an initial global state~$g_0$ of the form $g_0 = \< \emptyset, T_0>$,
where the initial global tree $T_0$ has one single node
labeled by $\< 0, \emptyset>$ ($0$ is the first location of $\M$).
The rest of the sequences of logical instants and global states is defined recursively 
as follows.

Given a global state $g_k = \< \gamma_k, T_k>$ at logical time $t_k \geq 0$, 
with $k \geq 0$, 
let $g'_k = \< \gamma'_k, T'_k> = g_k\downarrow_*$.
The next logical instant $t_{k+1}$ and global state $g_{k+1}$ are defined as follows.

\begin{description}
\item
If $\varepsilon$ unlocks $T'_k$ wrt $\gamma'_k$,
%Either$\< S'_k, 0>$ or $\< \gamma'_k, 0>$ enables $g'_k$, 
then $t_{k+1} = t_k$, 
and $g_{k+1} = \< \gamma_{k+1}, T_{k+1}>$ where
$T_{k+1} = T'_k\downarrow_{\varepsilon, \gamma'_k}$ 
and $\gamma_{k+1} = \gamma'_{k}$.
Note in particular that the signals are not reset in $\gamma_{k+1}$.
%and $g'_k \xrightarrow{S'_k, d} \circ \xrightarrow{\gamma'_k, d} g_{k+1}$.
%
\item 
Otherwise, $t_{k+1} > t_{k}$ is the date of the next logical event $\tau$,
%unlocking $T'_k$ wrt $\gamma'_k$,
which can be one of 
\begin{itemize}
\item[($i$)] $\mathsf{done}\, p$, 
\item[($ii$)] $\mathsf{step}\, p$, 
\item[($iii$)] $i \in \In$, 
\item[($iv$)] a global store $\alpha = \{ x \mapsto v \}$.
\end{itemize}
\noindent
Let $g_{k+1} = \< \gamma_{k+1}, T_{k+1}>$ where
$\gamma_{k+1}$ is obtained from $\alpha \circ\gamma'_{k}$
by resetting every signal assignment to $\false$
and $T_{k+1} = T'_k\downarrow_{\tau, \gamma'_k}$.
\end{description}

The execution depends on the behavior of the environment
but it is deterministic in the sense that the same behavior 
givens the same execution of $\M$.
Observational behavior can be characterized by the timed trace containing
the input symbols received with $\receive$, 
the global variables modified by the environment and 
the output symbols emitted with $\send$, 
each with the corresponding logical instant.

\section{Implementation Issues}

\subsection{Clock Services}
The clock services can be implemented using one
ordered queue of delays for each clock.

\subsection{Time Safety}
The above definition of execution is theoretical and assumes that 
the synchronous transition take zero delay.
In reality, we have to take care of the time needed to do these transitions.
Moreover, handling the events that define logical instants,
and the reorganization of the global tree, 
are assume instantaneous in Section~\ref{sec:sem}, 
we also need to take care of the time needed to perform these task in reality.
%we assume to have all the information needed
%on the arrival of input event from the environment.
%During a real execution, we only know the date of arrival of an event when it arrives.
Since there is no control on the environment these issues can not always be solved, 
let us discuss in this paragraph a best effort strategy to addresses them.

Let $g_k$ be a global state, reached at the logical instant $t_k$ 
(as defined in Section~\ref{sec:sem-time}).
%We assume that the logical time associated to $g_k$ coincide with $t_k$ 
%(at least this is the case for $i = 0$).
Let $\delta_k$ be the time needed to 
perform the synchronous transition and compute $g'_k = g_k \downarrow_*$
and let $\epsilon_k$ be 
the time needed for handling $i$ 
and making the synchronous transition from $g'_k$ to $g_{k+1}$.
%The time will be added to the next $\theta_{k+1}$.
For convenience, we let $\epsilon_{-1} = 0$.
Let $\theta_k = \delta_k + \epsilon_{k-1}$ for $k \geq 0$.

\noindent
Let us assume that the theoretical delay $d_k = t_{k+1} - t_{k}$,
as defined in Section~\ref{sec:sem-time},
corresponds to the arrival of an event $i$ (case ($iii$)).
If $d_k \geq \theta_k$, then time safety is ensured.
%difference between logical time and real time is $\epsilon_k$.
This can be depicted as follows, with the time flowing from left to right.

\quad
\(
\begin{array}{lclclcl}
g_k & & g'_k & & i & & g_{k+1}\;\mbox{ready}\\[1pt]\hline
t_k+\epsilon_{k-1} & \qquad & t_k + \theta_k & \qquad & t_k + d_k = t_{k+1} & \qquad & 
t_{k+1} + \epsilon_k\\
\end{array}
\)

\medskip
\noindent
If $d_k < \theta_k$, then there is a difference between 
logical time and real time that must be handled. %is now $\theta_k - d_k$.

\quad\(
\begin{array}{lclclcl}
g_k & & i & & g'_k & & g_{k+1}\;\mbox{ready}\\[1pt]\hline
t_k +\epsilon_{k-1} & \qquad & t_k + d_k = t_{k+1} & \qquad & t_k + \theta_k & 
\qquad & t_k +\theta_k+\epsilon_k\\
\end{array}
\)
%We keep track of the difference in a variable called $\Delta$.

\noindent
For instance, the difference $d_k - \theta_k$ 
can be retrieved from the delay of a $\await$ instruction 
occurring next to $g'_k$.
But we cannot guarantee that it is always possible.

\subsection{Static Analysis}
A strategy to predict statically time safety 
could be to use estimation of worst case execution time (WCET)
of the possible sequences of synchronous instruction in $\M$.
Note that these values depend on the execution platform.
Knowing on these durations, the analysis would then consist
in estimating whether the durations in the asynchronous $\wait$ 
instructions are compatible with the WCETs.
Moreover, one has to deal with the unpredictable timing for external events.
One approach could be to infer a linear constraint on these timings
for ensuring there compatibility with WCETs.
An alternative is to solve a 2 players safety game 
on the graph defined by the global states of $\M$, extended with the timing information.

The above approaches are similar to techniques used in the compilation 
of (X)Giotto into Ecode~\cite{Ghosal04xGiotto,HenzingerKirsch07Emachine}.
There are some differences however.
First, in Ecode, the analogous of the above synchronous instructions is
written in a conventional programming language like C, 
for which procedures for estimation of WCETs exist.
Second, all the timings in Giotto are expressed in milli-seconds, 
whereas timings can be expressed in multiple clocks in Antescofo.

Another interesting question in this setting is 
whether the structure of the intermediate code obtained from 
Antescofo programs is sufficiently simple in order to
avoid an exponential explosion in a time safety analysis.

%deterministic timing behavior.
Note that the execution of synchronous instructions following the ordering $\ll$
and the global execution scheme (decomposed into synchronous and asynchronous step)
permit to avoid race conditions and ensures determinism.

%%%%%%%%%%%%%%%%%%
%% REFERENCES
%%%%%%%%%%%%%%%%%%

\bibliographystyle{abbrv}
\bibliography{biblio}

\end{document}